\documentclass{PoS}

\title{Dynamical overlap fermion at fixed topology}

\ShortTitle{Dynamical overlap fermion at fixed topology}

\author{
  JLQCD collaboration:
  \speaker{S.~Hashimoto}$^{,a,b,}$\thanks{E-mail: shoji.hashimoto@kek.jp},
  S.~Aoki$^c$, 
  H.~Fukaya$^d$,
  K.~Kanaya$^c$,
  T.~Kaneko$^{a,b}$, 
  H.~Matsufuru$^a$,
  M.~Okamoto$^a$,
  T.~Onogi$^e$,
  N.~Yamada$^{a,b}$
  \vspace*{2mm}
  \\
  \llap{$^a$}
  High Energy Accelerator Research Organization (KEK),
  Tsukuba 305-0801, Japan
  \\
  \llap{$^b$}
  School of High Energy Accelerator Science,
  the Graduate University for Advanced Studies (Sokendai),
  Tsukuba 305-0801, Japan
  \\
  \llap{$^c$}
  Graduate School of Pure and Applied Sciences,
  University of Tsukuba, Tsukuba 305-8571, Japan
  \\
  \llap{$^d$}
  Theoretical Physics Laboratory, RIKEN, Wako 351-0198, Japan
  \\
  \llap{$^e$}
  Yukawa Institute for Theoretical Physics, 
  Kyoto University, Kyoto 606-8502, Japan
}

\abstract{
  We launched a project to perform
  dymanical fermion simulations using the overlap fermion
  formulation for sea quarks.
  In order to avoid the appearace of near-zero modes of the
  hermitian Wilson-Dirac operator $H_W$, we introduce a pair
  of extra Wilson fermions with a large negative mass term. 
  Crossing of the topological boundary is then strictly
  prohibited, and the topological charge is conserved during
  simulations.
  It makes the simulations substantially faster compared to
  the algorithms which allow the topology change.
  We discuss on the finite volume effects due to the fixed
  global topology.
}

\FullConference{XXIV International Symposium on Lattice Field Theory\\
                 July 23-28 2006\\
                 Tucson Arizona, US}

\begin{document}

\section{Introduction}
The JLQCD collaboration launched a new project to perform
dynamical fermion simulations with exact chiral symmetry
using the overlap fermion formulation for sea quarks.
It allows us to realize exact simulations of the chiral
symmetry breaking phenomena starting from the first
principles. 
With the exact chiral symmetry there is no fundamental
difficulty to simulate arbitrarily light quarks, which is
essential for the controlled chiral extrapolation
of many important physical quantities, such as the meson
decay constants, bag constants, form factors, and so on.
This talk is one of the first reports of our project
\cite{Kaneko_lat06,Yamada_lat06,Fukaya_lat06,Matsufuru_lat06},
which makes use of the new supercomputer system (Hitachi SR11000
and IBM BlueGene/L) installed at KEK in March 2006.

\section{Extra Wilson fermions}
The whole numerical difficulty of the overlap fermion comes
from the matrix sign function included in the definition of
the overlap-Dirac operator \cite{Neuberger:1998wv}
\begin{equation}
  \label{eq:Dov}
  D = m_0\left[1+\gamma_5\mathrm{sgn}(aH_W)\right],
\end{equation}
where $H_W$ stands for the hermitian Wilson-Dirac operator
$H_W=\gamma_5 D_W(-m_0)$ with large negative mass
$am_0=1.6$.
One usually employs polynomial or rational function to
approximate the sign function, which becomes increasingly
difficult when there are many near-zero eigenmodes of $H_W$.
Furthermore, the molecular dynamics Hamiltonian in the HMC
simulation involves discontinuity when the near-zero mode
pass through zero.
Such discontinuity may be treated exactly using the
so-called ``reflection/refraction'' trick introduced by
Fodor, Katz, and Szabo \cite{Fodor:2003bh}, 
but its numerical cost scales at least as $\sim V^2$ as the
lattice volume $V$ increases, and the simulation on
reasonably large lattices, {\it e.g.} $16^3\times 32$ as
used in our work, becomes prohibitively costly. 

The near-zero modes of $H_W$ appear on rough gauge
configurations.
In fact, the corresponding eigenvector develops a local
lump, {\it i.e.} an exponentially localized wave function
with support length as small as a few lattice sites
\cite{Berruto:2000fx,Golterman:2003qe,Golterman:2005fe,%
Yamada_lat06}. 
The net topological charge of the gauge field as defined
through the index of $D$ can change its value only through
such rough gauge configurations.
It implies that the problem of near-zero modes is one of the
lattice artifacts disappearing in the continuum limit.
A relevant question is, then, whether such artifacts can be
removed at finite lattice spacings by choosing lattice
action appropriately.

In fact, one may construct a class of lattice actions for
which the exact zero mode is prohibited by introducing extra
Wilson fermions with a large negative mass $-m_0$
\cite{Izubuchi:2002pq,Vranas:2006zk,Fukaya:2006vs}.
For the two-flavor case it produces a determinant factor
$\det H_W^2$ which suppresses near-zero modes of $H_W$ and
strictly prohibits the appearance of exact zero modes.
In the continuum limit, the extra fermions become irrelevant
as they have a mass of order of lattice cutoff.

We also add a pair of twisted mass pseudo-fermions in order
to cancel the unwanted effects of the extra fermions for
higher modes.
Namely, we introduce a determinant factor
$\det[H_W^2/(H_W^2+\mu^2)]$ with $\mu$ a twisted mass.
Since the effect on higher modes is largely cancelled in the
ratio, the finite renormalization of the parameters can be
minimized.
For instance, the $\beta$-shift, necessary change of the
$\beta$ value to keep the same lattice spacing, is 0.786(4)
without psuedo-fermions, but is reduced to 0.050(1) with the
pseudo-fermions with twisted mass $\mu=0.2$
\cite{Fukaya:2006vs}.

The numerical evidences of the effect of the extra Wilson
fermions are extensively studied in the quenched
approximation (in the sense that the dynamical overlap
fermion is not included) and presented in
\cite{Fukaya:2006vs,Yamada_lat06}.
To be short, the problematic near-zero modes essentially
disappear while keeping the $\beta$-shift minimal.

\section{Dynamical overlap fermions}
Our methods for simulating dynamical overlap fermions are
rather standard except for the introduction of extra Wilson
fermions as discussed above.

We approximate the sign function using the rational function
with Zolotarev coefficients.
Lowest-lying modes below some threshold are treated exactly.
With (typically) 10 poles for the rational function we keep
the precision of the sign function to the level of
$10^{-(7-8)}$.
For the solver of the overlap-Dirac operator we use the
conjugate gradient algorithm with relaxed precision
\cite{Cundy:2004pz}.
In the inner loop for the rational function calculation, we
use the multi-shift CG algorithm.
(More recently, we tested a five dimensional implementation
of the overlap solver \cite{Edwards:2005an} and found it is
about $\times$4 faster. 
The details are discussed in \cite{Matsufuru_lat06}.)

In the molecular dynamics evolution we introduce the
Hasenbusch's mass preconditioner \cite{Hasenbusch:2001ne}
with multi-time steps:
(1) inner-most loop containing gauge and Wilson fermion
forces, (2) middle loop treating the mass preconditioner
(Hasenbusch) for the overlap fermion, and (3) outer loop
with preconditioned overlap fermion.

The numerical simulations described in this talk have been
done on a $16^3\times 32$ lattice.
We use the Iwasaki gauge action.
The twisted mass $\mu$ of the extra psuedo-fermions is fixed
to 0.2. 
The $\beta$ value is chosen such that the lattice spacing
determined through $r_0$ becomes 0.11$-$0.12~fm.
As a result we have two series of runs at $\beta$ = 2.35
($a$ = 0.11~fm) and at $\beta$ = 2.30 ($a$ = 0.12~fm).
In both cases we performed simulations at six values of
quark masses corresponding to $m_s/6-m_s$.
Our main data are those of $\beta=2.30$, whose analysis is
discussed in more detail in \cite{Kaneko_lat06}, while the
$\beta=2.35$ run has been performed for parameter tuning.
At $\beta=2.35$, we also carried out a test run in the
$\epsilon$-regime: quark mass is so small (in our test run,
$\sim$ 2~MeV) that $m_\pi L\ll 1$.
Its analysis is described in \cite{Fukaya_lat06}.
The topological charge is fixed to 0 for all runs.

For a comparison, we also implemented the
reflection/refraction steps and carried out a test run at
$\beta=2.45$ with a relatively heavy sea quark mass 
($\sim m_s$). 
In this case we switched off the extra Wilson fermions.

\begin{figure}[tb]
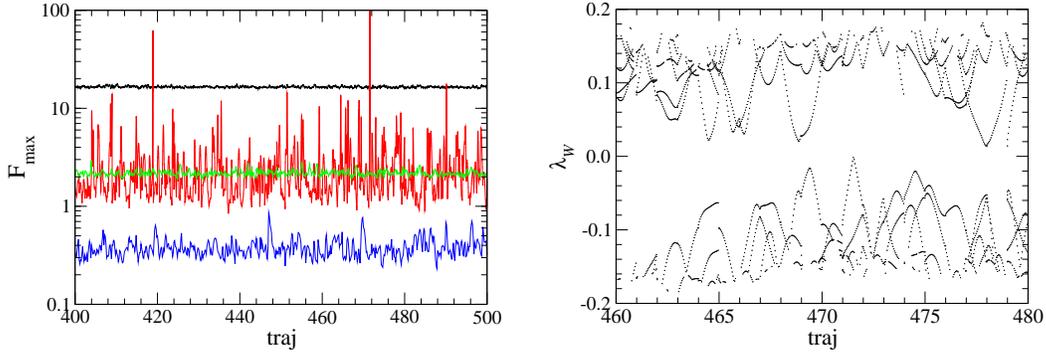

  \centering
  \includegraphics[width=6.6cm,clip=true]{figure/force.eps}
  \hspace*{4mm}
  \includegraphics[width=6.6cm,clip=true]{figure/mdeigen_b2.35_m110.eps}
  \caption{
    Left: History of maximum molecular dynamics force for
    gauge (black), extra Wilson fermion (red),
    preconditioner overlap (green), and preconditioned
    overlap (blue). 
    Right: History of lowest-lying eigenvalues of $H_W$.
    Data at $\beta=2.35$ and $am$ = 0.110.
}
  \label{fig:history}
\end{figure}

Here we discuss the effect of the extra Wilson fermion in
the molecular dynamics evolution.
In Figure~\ref{fig:history} we plot the HMC history of the
maximum molecular dynamics force (left) and the lowest-lying
eigenvalues of $H_W$ (right).
The force term is calculated at each step of the molecular
dynamics evolution and its maximum value
over the whole lattice sites is plotted on the left panel.
We clearly see a hierarchy ``gauge'' $\gg$
``preconditioner'' $\gg$ ``preconditioned'', justifying our
choice of the multi-time steps.
Among other forces, that from the extra Wilson fermions
(plus psuedo-fermions) has the largest fluctuation, and
sometimes becomes larger than the gauge force.
This is the reason that we include this force calculation in
the inner-most loop together with the gauge force.
The right panel shows how the extra Wilson fermions repel
the near-zero eigenvalues.
There are quite a few cases where the lowest eigenvalue
tries to accross zero, but it is bounced back by the
repulsive force.
As a result, the net topological charge never changes its
value during the molecular dynamics evolutions.

\begin{figure}[tb]
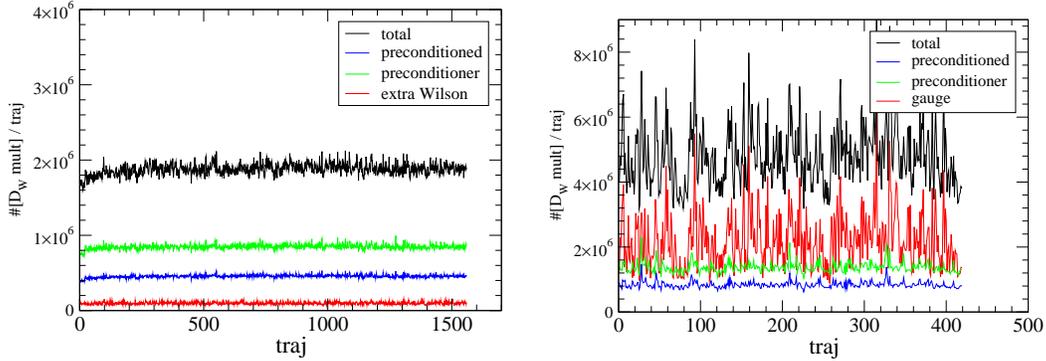

  \centering
  \includegraphics[width=6.6cm,clip=true]{figure/mdcost.eps}
  \hspace*{4mm}
  \includegraphics[width=6.6cm,clip=true]{figure/refcost.eps}
  \caption{
    Numerical cost per trajectory measured by the number of
    the Wilson-Dirac operator multiplication.
    Left: the cost with the extra Wilson fermions.
    Right: the cost without the extra Wilson fermions but
    with the reflection/refraction steps.
    In addition to the total cost (black), its breakdown to
    preconditiond overlap (blue), preconditioner overlap
    (green) and extra Wilson (red) are shown.
    On the right panel, the red shows the cost for gauge
    steps including the reflection/refraction.
    Data at $\beta$ = 2.35 and $am=0.110$ (left) and at
    $\beta$ = 2.45 and $am=0.090$ (right).
  }
  \label{fig:cost}
\end{figure}

Although the extra Wilson fermions are put in the inner-most
loop and they require a solver of the Wilson fermion at
every steps, their numerical cost is not substantial.
Figure~\ref{fig:cost} shows the breakdown of the numerical
cost per HMC trajectory.
As plotted in the left panel, we find that the cost for the
extra-Wilson fermion is several times smaller than that for
the overlap fermion forces and is only about 5\% of the
total cost. 
The right panel, on the other hand, shows the cost with the
reflection/refraction steps.
We can see that about a half (or more) of the total
numerical cost is spent for the gauge steps, which contain
the monitoring of the lowest-lying eigenvalues at the every
steps and the reflection/refraction when they tend to across
zero.
Once the reflection/refraction steps are entered, two
inversions of the overlap operator are required, and this is
the reason why the cost is fluctuating very much.
Not just the fluctuation but the average magnitude is more
than a factor of two larger than the case with the extra
Wilson fermions (and thus without reflection/refraction).
Furthermore, we expect that the cost for
reflection/refraction increases as $V^2$ as the lattice
volume $V$ increases and becomes a limiting factor in larger 
volume simulations.
This is the reason that we do not take this option in our
main runs.

\section{Topology issues}
The advantage of the fixed topology simulation with the
extra Wilson fermion is not just its lower numerical cost,
but it suppresses a certain class of discretization effects
coming from dislocations.
An obvious disadvantage, on the other hand, is the fixed
topological charge during the HMC simulations, which implies
that the correct $\theta=0$ vacuum can not be sampled.

In this project we rely on the cluster decomposition
property of the local field theory, which suggests that the
global topology is irrelevant for local physics as far as
the volume is large enough.
Our expectation is that the topological fluctuations
are active in local areas of the lattice, even though the
global topology is fixed.
The topological susceptibility $\chi_t$ controls the
frequency of the local fluctuations.
In the instanton picture of the QCD vacuum, the local
topological fluctuation could occur through
an instanton-anti-instanton pair creation.
Once an instanton, which is a local object, is created, it
may move around the lattice until it meets an anti-instanton
to pair annihilate.
The local topological fluctuation should then manifest
itself in the space-time distribution of associated local
zero modes, which would become localized near-zero modes
when measured on the whole lattice.

\begin{figure}[tb]
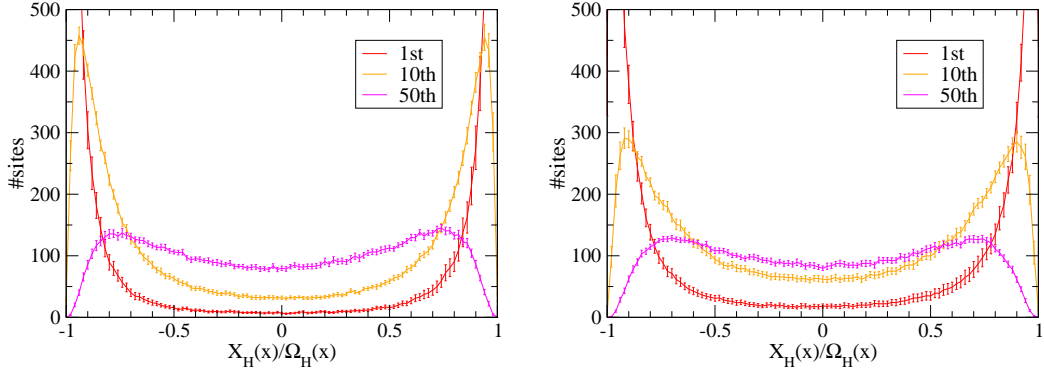

  \centering
  \includegraphics[width=6.6cm,clip=true]{figure/correlation.eps}
  \hspace*{4mm}
  \includegraphics[width=6.6cm,clip=true]{figure/correlation_dyn_m110.eps}
  \caption{
    Correlation of the density
    $\Omega_i(x)=\psi_i^\dagger(x)\psi_i(x)$ and chiral
    density $X_i(x)=\psi_i^\dagger(x)\gamma_5\psi_i(x)$.
    Results for the lowest, 10th, and 50th eigenvalues are
    shown after averaging over 50 gauge configurations.
    Data are for a quenched lattice at $\beta=2.37$ (left)
    and for a dynamical run at $\beta=2.35$, $am=0.110$
    (right). 
  }
  \label{fig:corr}
\end{figure}

As a demonstration we look at the correlation between the
density $\Omega_i(x)=\psi_i^\dagger(x)\psi_i(x)$ and chiral
density $X_i(x)=\psi_i^\dagger(x)\gamma_5\psi_i(x)$ of the
low-lying eigenvectors $\psi_i(x)$ of the overlap-Dirac
operator.
Here, $i$ labels different eigenvectors.
As found in the previous works
\cite{Horvath:2001ir,DeGrand:2001pj,Blum:2001qg}, the
distribution of the ratio $X_i(x)/\Omega_i(x)$ peaks at 
$\pm 1$ when the local lumps of the eigenmode $\psi_i(x)$ are
predominantly chiral.
(We put a cut on $\Omega_i(x)$ such that only the points
satisfying $\Omega_i(x)>1/V$ are counted.)
Figure~\ref{fig:corr} shows our results for quenched (left)
and dynamical (right) lattices.
The global topology is fixed to 0 on both lattices.
The double peak structure is clearly seen especially for the
lowest-lying eigenmode on both lattices.
This result supports our expectation that there are local
topological lumps even in the fixed topology simulations. 

Hadron masses calculated at a fixed topological charge
may deviate from their values in the $\theta=0$ vacuum.
Within the framework of chiral lagrangian it is possible to
calculate such effects \cite{Brower:2003yx}.
The difference is predicted as
$M''(0)/(2\langle Q^2\rangle) \left(1-Q^2/\langle Q^2\rangle\right)$,
where $M''(0)$ is a second derivative of the relevant hadron
mass with respect to $\theta$.
Since $\langle Q^2\rangle=\chi_t V$ appears in the
denominator, the deviation behaves as $1/V$.
Although the effect becomes larger for smaller quark masses
(for small enough $m$ one expects $\chi_t=m\Sigma/N_f$), 
the size is $O(1\%)$ on our smallest quark mass, for which
the pion mass is around 300~MeV.

As a first test we compare the pion mass calculated in the
quenched approximation at several fixed topological charges.
On a $16^3\times 32$ lattice at $\beta$ = 2.37 ($a\simeq$
0.125~fm), we have carried out quenched simulations with the 
extra Wilson fermion, starting from gauge configurations
with definite topological charges $Q$ = 0, 2, 3 and 6.
For each $Q$ we accumulate 100 gauge configurations
each separated by 200 HMC trajectories and measure pion
masses.
From the results in the quenched continuum limit
\cite{DelDebbio:2004ns} we expect the average
topological charge squared on our lattice to be $\langle
Q^2\rangle$ = 33(3), which leads to an expectation of fairly
small effect $\sim O(0.2\%)$ at the tree level of chiral
effective theory.

\begin{figure}[tb]
  \centering
  \includegraphics[width=6.6cm,clip=true]{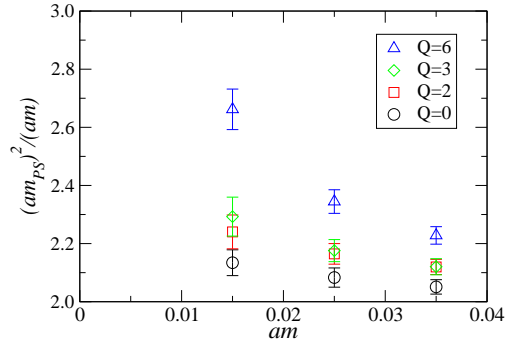}
  \caption{
    $m_{PS}^2/m_q$ for different topological charges $Q$ =
    0, 2, 3, 6.
  }
  \label{fig:mPS2}
\end{figure}

The results for $(am_{PS})^2/(am_q)$ are plotted in
Figure~\ref{fig:mPS2}. 
The data show much larger dependence on the topological
charge.
A similar dependence is also recently reported in
\cite{Galletly:2006hq}.
This result is somewhat unexpected, and probably due to
the quenched artifact.
In the quenched chiral perturbation theory, the one-loop
correction to the pion mass comes from a tadpole diagram
with a hairpin insertion.
Because of the double pole of singlet pion, the diagram has
an infrared divergence in the massless pion limit.
Since the hairpin vertex picks up the singlet mass squared
$m_0^2$, which is proportional to $\chi_t$, the one-loop
diagram leads to the quenched chiral logarithm proportional
to the global topological charge $Q^2$.
This is one of the pathologies of quenching.

In order to investigate the topology and volume dependence
of the hadron masses, we are plannning to carry out
simulations at diffrent $Q$ and $V$.

\section{Summary}
We reported the current status of our project of dynamical
overlap fermion simulation.
Although the calculation of the overlap-Dirac operator is so
costly, it turned out that its dynamical simulation is
feasible if we fix the global topological charge to suppress
the near-zero modes of $H_W$.
In fact, we are able to simulate a $16^3\times 32$ lattice
with quark mass as small as $m_s/6$.

Study of the effects of fixed topological charge is
underway.
We found that there are local chiral (and thus topological)
lumps in the gauge configurations.
More study would be necessary to fully understand the
effects on physical quantities such as the hadron masses. 

\vspace*{4mm}
Numerical simulations are performed on Hitachi SR11000 and
IBM System Blue Gene Solution at High Energy Accelerator Research
Organization (KEK) under a support of its Large Scale
Simulation Program (No. 06-13).
This work is supported in part by the Grant-in-Aid of the
Ministry of Education 
(No. 13135204, 13135213, 15540251, 16740156, 17340066, 17740171, 18034011, 18340075, 18740167).

\end{document}